\documentclass[a4paper,11pt]{article}
\pdfoutput=1 

\usepackage{jinstpub} 

\title{Medipix3 proton and carbon ion measurements across full energy ranges and at clinical flux rates in MedAustron IR1}

\author[a,b,1]{N. J. S.~Bal,\note{Corresponding author.}}
\author[c]{C. S.~Schmitzer,}
\author[c]{A.~De Franco,}
\author[c]{S.~Enke}

\affiliation[a]{Nikhef,\\Science Park 105, 1098 XG Amsterdam, Nederland}
\affiliation[b]{CERN,\\CH-1211 Geneva 23, Switzerland}
\affiliation[c]{EBG GmbH MedAustron,\\Marie-Curie-Stra{\ss}e 5, 2700 Wiener Neustadt, \"{O}sterreich}

\emailAdd{navritb@nikhef.nl}

\abstract{
The Medipix3, a hybrid pixel detector with a silicon sensor, has been evaluated as a beam instrumentation device with proton and carbon ion measurements in the non-clinical research room (IR1) of MedAustron Ion Therapy Center. Protons energies are varied from 62.4 to 800 MeV with $10^{4}$ to $10^{8}$ protons per second impinging on the detector surface. For carbon ions, energies are varied from 120 to 400 MeV/amu with $10^{7}$ to $10^{8}$ carbon ions per second. Measurements include simultaneous high resolution, beam profile and beam intensity with various beam parameters at up to 1000 FPS (frames per second), count rate linearity and an assessment of radiation damage after the measurement day using an x-ray tube to provide a homogeneous radiation measurement. The count rate linearity is found to be linear within the uncertainties (dominated by accelerator related sources due to special setup) for the measurements without degraders. Various frequency components are identified within the beam intensity over time firstly including 49.98 Hz with standard deviation, $\sigma=0.29$, secondly 30.55 Hz $\sigma=0.55$ and thirdly 252.51 Hz $\sigma=0.83$. A direct correlation between the number of zero counting and noisy pixels is observed in the measurements with the highest flux. No conclusive evidence of long term radiation damage was found as a result of these measurements over one day.
}

\keywords{Instrumentation for hadron therapy, 
Beam-line instrumentation (beam position and profile monitors, beam-intensity monitors, bunch length monitors), Radiation damage to detector materials (solid state), Instrumentation for particle-beam therapy}

\proceeding{22$^{\text{nd}}$ International Workshop on Radiation Imaging Detectors\\
  27 June 2021 to 1 July 2021\\
  Ghent, Belgium (online)}

\begin{document}
\maketitle
\flushbottom



\section{Introduction}

The Medipix3 chip is a hybrid pixel detector from the family of chips developed by the Medipix group at CERN \cite{ballabriga}  and have found many applications from electron microscopy \cite{PATERSON2020112917} to spectral x-ray microCT \cite{ronaldson2012quantitative}. The Medipix3 manual \cite{medipix3rx-manual-v1.4} contains detailed information on how the chip works.

Previous count rate linearity measurements of the Medipix3 with x-rays at a synchrotron \cite{1748-0221-9-04-C04028}, show the count rate to be linear up to the order of $10^{8}$ photons / mm$^2$ / second. This is significantly more than the maximum expected particle flux rate in these measurements of approximately $10^{4}$ to $10^{6}$ particles / mm$^2$ / second. Therefore, assuming the front-end behaves similarly enough, the count rate linearity is expected to be comparable between proton, carbon ion and x-ray measurements. Given that the Medipix3 is designed for relatively low energy x-ray detection ($<$100 keV) and we are using 60+ MeV particles, this assumption is significant and has been verified with the measurements presented in this work.

This work follows a first measurement with protons in a clinical environment at the Clatterbridge Cancer Centre (CCC), UK \cite{Yap:2019zrh}. It was demonstrated that the count rate was linear within the uncertainties from the beam variation. The CCC beam current measurements have large uncertainties due to recording the beam current only once per measurement while it was varying in the order of 10\%.



\section{Experimental details}

Measurements took place in IR1 (irradiation room 1) which is a non-clinical research room, this is not a standard treatment room but solely for research purposes. The active detector area was 28 $\times$ 28 mm$^2$, consisting of 4 Medipix3 chips in a 2 $\times$ 2 grid with a gap of approximately 220 $\upmu$m between each chip. The sensor material was high resistivity silicon, with a thickness of 500 $\upmu$m and pixel pitch (spatial resolution) of 55 $\upmu$m. The Medipix3 was primarily designed for x-rays and electron detection, however it detects any ionising radiation depositing $>5\:$keV within a single pixel. The possible frame rates are 0---2000 FPS for 12 bit pixel counter depth (0 to 4095 counts per pixel per frame) in Continuous R/W (Read/Write) mode (0 dead time). Other pixel counter depths and readout modes have different frame rate maximums which depend on the readout frequency of the chip, this is running at 200 MHz. Higher frame rates are possible with lower counter bit depths, a frame rate of up to 24 kHz is possible with a 1 bit counter depth; this mode results in a binary hit map. The detector frame rates used in this work were 50, 100, 1000 FPS in Continuous R/W mode (0 dead time). Ideally one would use the maximum frame rate of 2000 FPS, however the readout computer used was not fast enough to reliably readout the data and so an upper limit of 1000 FPS was set. 50 and 100 FPS were used when the flux rate was low and the higher frame rate was not necessary in order to save disk space.

\begin{figure}[b]
    \centering
    \includegraphics[width=0.5\textwidth]{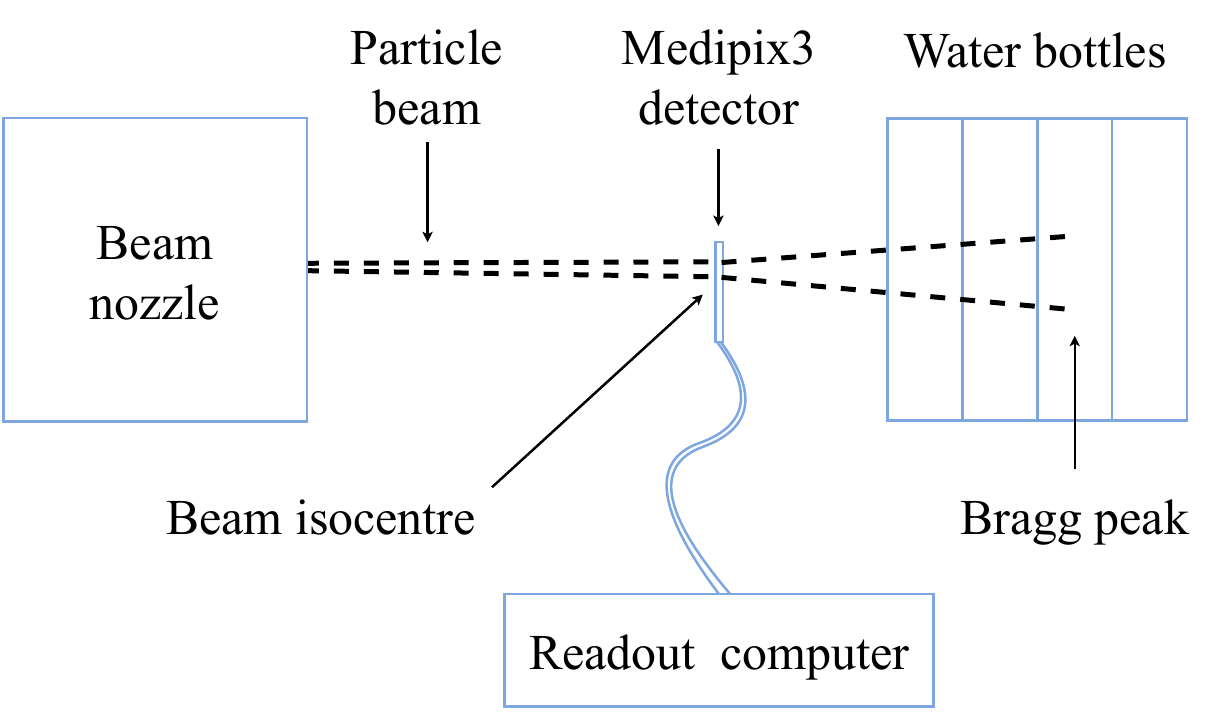}
    \caption{The equipment layout from left to right included a beam nozzle from which the particle beam is emitted, the detector was positioned in the beam isocentre using alignment line lasers. The beam isocentre is the central axis in space aligned to lasers where the centre of the target (within the patient) is positioned. The Medipix3 detector (300 $\times$ 100 $\times$ 100 mm$^{3}$) contains a detector assembly where the radiation is measured and SPIDR v3.5 (Speedy PIxel Detector Readout) system. The SPIDR was connected to a readout computer via a 10 Gbit/s optical fibre where the data was stored. Finally the beam enters the water bottles of approximately 2 m depth so the Bragg peak was always within the bottles, these were used for shielding the robotic arm.} 
    \label{fig:equipment}
\end{figure}

The high level experimental setup overview is shown in figure \ref{fig:equipment}.

Due to driver limitations at the time of measurements, it was only possible to operate with SPM (Single Pixel Mode) and not CSM (Charge Summing Mode). This means that the pixels were individually counting, the charge was not summed over a 2x2 grid. At low flux, it is very likely that CSM would improve the PSF (Point Spread Function). The benefit would be maximised when the charge cloud covered an average of 4 pixels in a 2x2 grid. CSM requires inter-pixel communication which takes time and therefore for the same front-end configuration, the count rate linearity for CSM degrades at lower count rates than SPM \cite{1748-0221-9-04-C04028}.

Only the first threshold was used for all measurements in this work. Table \ref{tab:detector-frontend-dacs} shows the relevant front-end DACs (Digital to Analogue Converter), none were changed during the measurements. With this configuration, the threshold was set at approximately 5 keV. The threshold was left at the minimum but just above noise value which negatively impacts the PSF compared to setting the threshold at a much higher level given the energy deposition of the particle species and energies investigated in this work.
\begin{table}[tb]
\centering
\caption{ Relevant front-end DACs used and kept constant during all measurements. DAC\_DiscL has one value per chip in order. All other rows are common between all four chips. }
\smallskip
\begin{tabular}{|c|c|c|}
\hline
            & Value          & Units \\
\hline
Threshold 0 & 42             & DAC   \\
Gain Mode   & High           & N/A   \\
DAC\_DiscL  & 71, 77, 75, 69 & DAC   \\
FBK         & 173            & DAC   \\
GND         & 121            & DAC   \\
IKrum       & 10             & DAC   \\
Preamp      & 150            & DAC   \\
Shaper      & 150            & DAC  \\ \hline
\end{tabular}
\label{tab:detector-frontend-dacs}
\end{table}

This detector system is highly optimised for x-ray detection between 4---30 keV. 500 $\upmu$m of silicon limits the energy range up to approximately 30 keV because it is increasingly transparent as x-ray energy increases, see figure~9 \cite{esrf}. The particle flux detected was $10^3$---$10^9$ particles per second over the active area of the detector. The lower limit of detection is single particles. The upper limit for protons and carbon ions is a topic of investigation in this work. The maximum flux that the synchrotron can deliver is $10^{10}$ protons per second.

The proton energies used were 62.4, 148, 252, 800 MeV, and for carbon ions, 120, 260, 400 MeV/amu (atomic mass unit) ions were used. The motivation for using 800 MeV protons is for proton CT which is being investigated at MedAustron in order to measure the proton relative stopping power (RSP) with respect to water. The current method uses x-ray CT which gives the x-ray attenuation in HU (Hounsfield unit) which are then empirically mapped to relative proton stopping power, see Wayne D Newhauser et al 2008 Phys. Med. Biol. 53 2327 Figure 3: "Relative linear stopping power for protons (dE/dx|$^x_w$) as a function of the scaled Hounsfield unit value ($H_x$, in units of $HU_{sc}$) in kVCT, where x denotes a material of interest and w denotes water." \cite{protonCT}. This figure contains one line plot consisting of three fitted straight lines with different slopes and intercepts with significant outliers; it is not a simple and clean linear relationship. This is one of the most important calibrations in this context because it feeds into every single dose distribution calculation of every patient. This conversion introduces one of the main sources of uncertainty in proton therapy treatment planning \footnote{This is common knowledge in the Medical Physics community} and is an area of active research \cite{Ainsley2014}.

\begin{figure}[tb]
    \centering
    \includegraphics[width=0.4\textwidth,keepaspectratio]{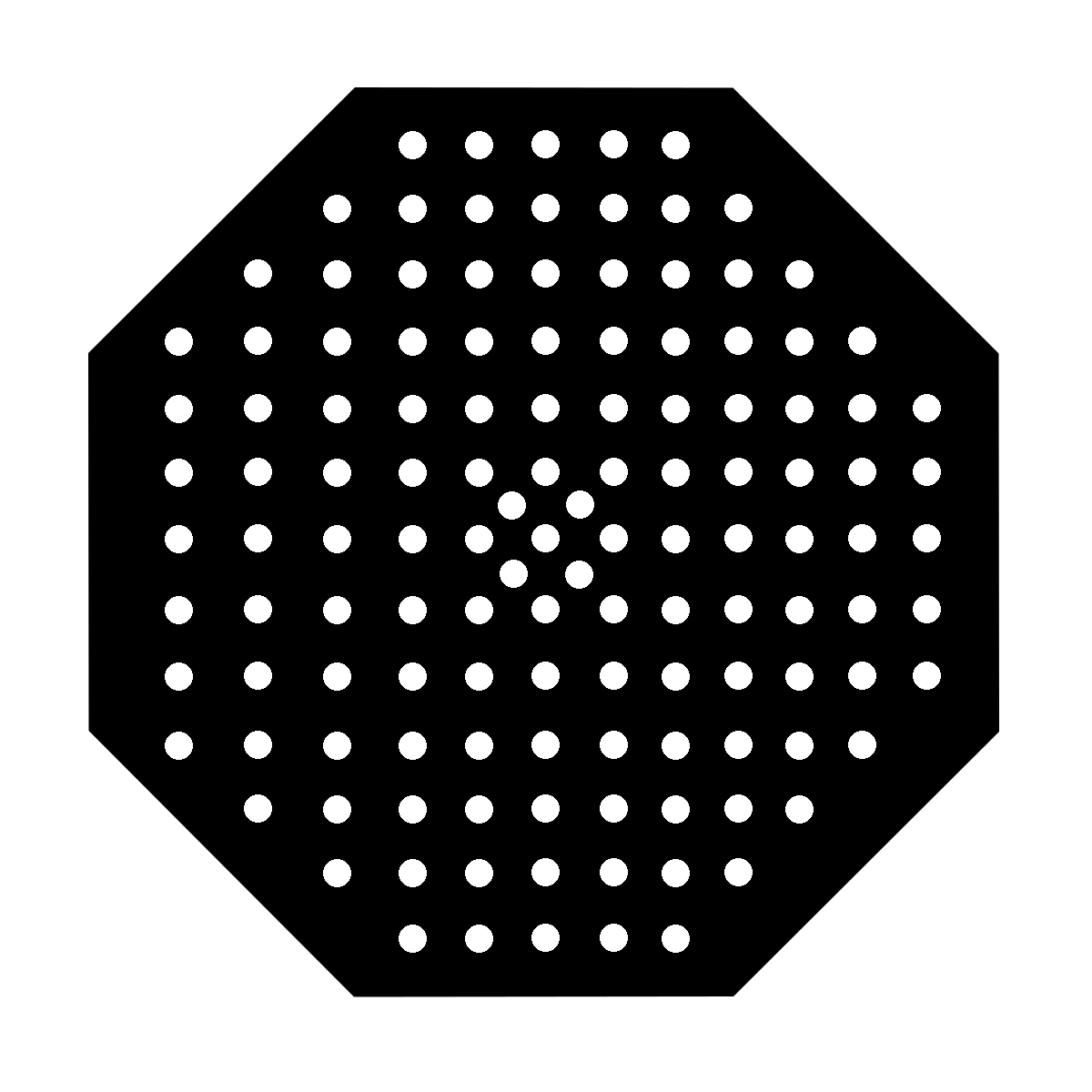}
    \caption{ An example diagram of a pepper pot degrader plate where black is steel, the white circles are through-holes and the plate has a thickness in the axis of the diagram. There is a higher hole density in the centre because the beams are approximately 2D normal distributions and the beam primary axis is aligned with the centre of the degrader plates. The plate is in the order of 30 cm diameter and the holes are in the order between mm and cm.}
    \label{fig:20210419-MedAustron-DegraderDiagram}
\end{figure}

The degrader plates used in this work were: 10, 20, 50 \& 100 \%. The degrader percentage is the nominal hole to surface ratio, ideally this would translate to a given percentage of incoming particles being transmitted relative to the total incoming number, while minimally affecting the energy. In reality the effective transmission is different. Some of the reasons behind the discrepancy are the self induced small charge forces and the slightly different accelerator optics used to compensate for such effects. As these effects vary with beam species and energy, a priori the flux reduction might differ from the nominal value according to the beam delivered. The quantitative effectiveness of the flux reduction of each degrader, which is a function of energy and species, is obtained with a series of monitors at each stage of the accelerator and in the irradiation room also. A degrader plate is a passive device, it is simply a steel plate with various numbers of holes of various sizes, this is referred to as a `pepper pot' design. Ideally this produces an identical distribution of particles as the incoming beam. An example diagram of a pepper-pot degrader can be seen in Figure \ref{fig:20210419-MedAustron-DegraderDiagram}, further technical details such as the specific geometry of the MedAustron degraders are not publicly available. After the degrader plates, the beam is shaped with beam optics which focus it at the isocentre and therefore are still compatible with pencil beam scanning. The degraders are one of the possible ways to regulate the particle fluence. Some measurements used degrader 100\% (no degrader) and others used degrader plates (10 \%, 20 \% and 50\%).

\section{Results and discussion}
\subsection{Count rate linearity}

Accuracy and precision are both hard to determine because there were no more accurate or precise detectors or methods available to measure flux at this range.

For example, gas ionisation chambers which are commonly used in medical accelerators, saturate at relatively low count rates due to the relatively low charge carrier density in gas. They also suffer from not being able to detect low count rates because the signal induced by single particles is not detectable. Solid state detectors have a very high charge carrier density compared to gas which means that the saturation limit is much higher. Gas based detectors benefit from being radiation hard because there is no crystalline structure to damage and the detection medium can be easily replaced.

Count rate linearity is important for accelerator calibration purposes because it makes calibration much simpler with lower associated fitting errors and maintains count rate precision over the entire range. Typically count rate linearity is assessed over either chip or pixel level. The linearity is assessed for all 4 chips for the simplicity of analysis and simplicity of the expected relationship between expected and measured count rate. In addition, the way these measurements were performed is the intended use for the detector.

The RayStation TPS (treatment planning system) from RaySearch Laboratories AB \cite{RayStationTPS} was used to request $196$ spots in order to irradiate the whole detector with a relatively homogeneous field. The FWHM (full width half maximum) spot size varied from 7 to 21 mm for protons and from 6.5 to 9.5 mm for carbon beams \cite{CarbonCommMA2019}.
The requested spot weight was varied between $5 \times 10^{6}$ and $1 \times 10^{9}$ particles.

The summed count on the Medipix3 is the sum of every count on every frame. It is expected that this should be very linear if the relative uncertainties on the expected proton fluence are sufficiently low. This is measured over the largest possible range of proton fluence, using degrader plates 10, 20 and 50 for the lowest three fluences. The lowest three fluences were chosen due to limited measurement time and the desire to probe the lower count rate region where the detector was expected to have a better count rate linearity. Spot weights were determined based on the available pre-configured options in the control system, they are clinically relevant and are expected to be well within reasonable limits of all relevant systems.

Uncertainties in this measurement are described as follows. Firstly, not all the spots are entirely on the detector, given the relatively large FWHM of the beam at isocentre; some of the beam will not hit the detector. This effect could be quantified with information from the TPS. A brief analysis of a `single' spot over 20 ms shows that this is likely not the case given the decreased linearity.

Secondly, it is observed at ultra low proton counts that 62.4 MeV protons produce a cluster of approximately 4---5 pixels. Ideally it would count once per proton or carbon ion. The detector will therefore count 4---5 times per 62.4 MeV proton as a result of this effect. As the energy increases, the cluster size decreases as less charge is deposited over the sensor depth which causes fewer pixels to count a hit. As the intensity increases, the events overlap within a single frame (minimum of 1 ms) and so cluster size per particle cannot be determined. This happens at the lowest clinical flux rates as they are high relative to the frame time. If one were able to increase the frame rate to infinity then the cluster size for high intensity measurements should not vary compared to low intensity measurements.

Cluster analysis could be done of the ultra low flux measurements to quantify this effect. This analysis was not done because the outcome would only give information about cluster sizes, their average, minimum and maximum values and would be affected by background measurements from both radioactive decay products from the activated sensor and surroundings and much less so, cosmic rays. It was therefore not expected to yield useful information as the focus is on count rate linearity rather than an absolute particle count. In order to measure an absolute particle count accurately, single cluster analysis would be necessary and is not achievable with this frame rate and flux.


Finally, there are shot-to-shot variations in the actual particle flux, as seen in table \ref{tab:p800-shot-to-shot-variation} for 800 MeV protons. Other proton energies and carbon ion energies were not scanned over with different degraders due to time constraints. It is assumed that this is due to variations in the extraction process or beam current from the synchrotron.

\begin{figure}[tb]
    \centering
    \includegraphics[width=0.7\textwidth,keepaspectratio]{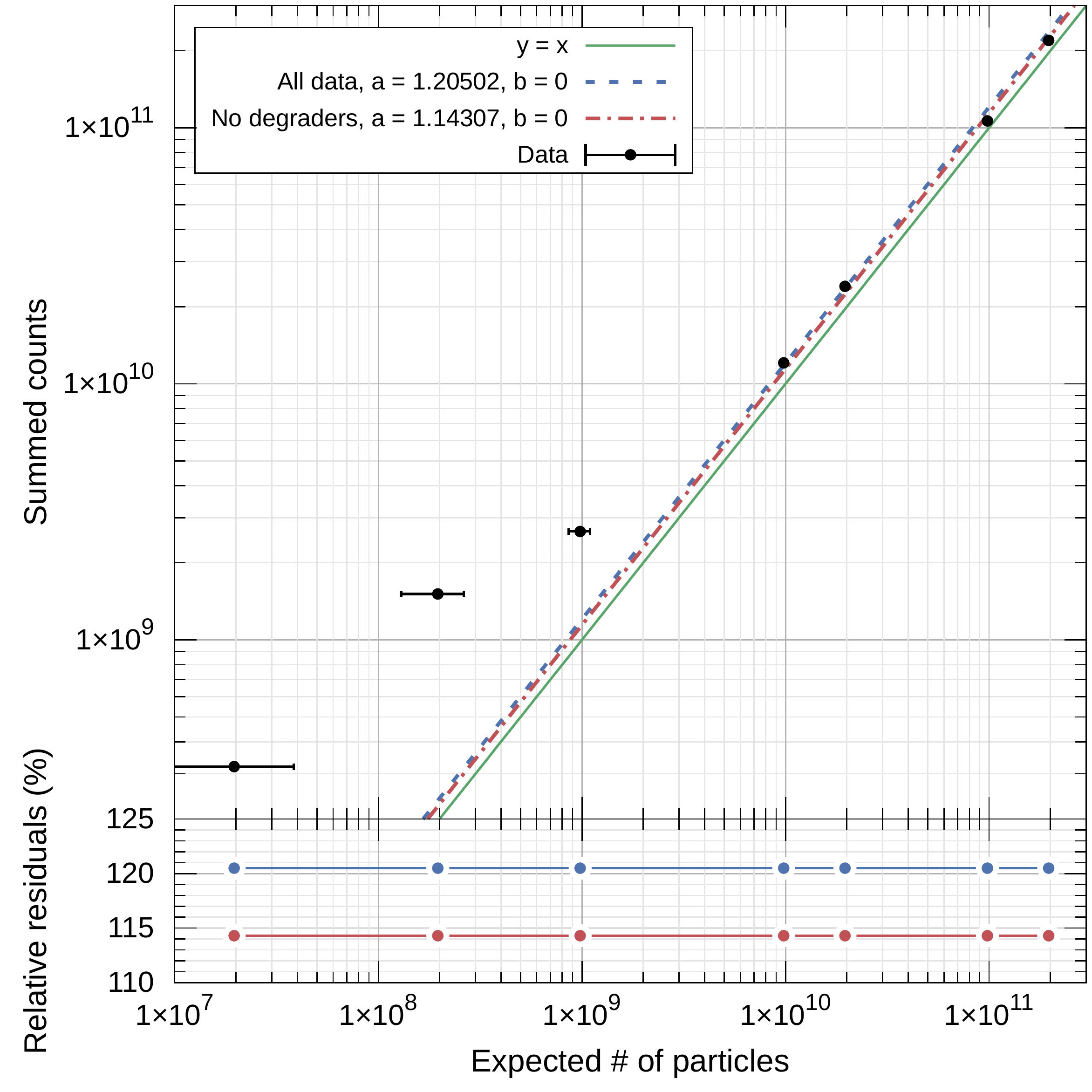}
    \caption{ The count rate linearity of all pixels integrated over all frames for the measurement using 62.4 MeV protons. Fits are shown both with degraders (a = 1.20502, b is fixed at zero) and without degrader measurements (a = 1.14307, b is fixed at zero) with the associated relative percentage residuals from the fits. The reduced chi squared value for the fit with all data points is $\chi_\nu^{2}$ = 91 and for the fit with no degrader measurements is $\chi_\nu^{2}$ = 4.6. The intercept parameter (b) is fixed at zero because zero counts are measured when the accelerator is not delivering any particles and counts from sensor activation are negligible at a few hundred counts per second. The degrader measurements are the first three data points from the left and are not included due to large systematic uncertainties as described in the text and seen in \ref{fig:p800-count-rate-linearity-all-data}. The fit algorithm was the non-linear least-squares fit (NLLS) Marquardt-Levenberg.}
    \label{fig:p62-count-rate-linearity}
\end{figure}  
    

Figure \ref{fig:p62-count-rate-linearity} contains the data points with uncertainties in the expected number of particles only with two fits; one with the degrader measurements and the other without. y = x is plotted as a reference to the naive expected relationship between the summed counts and the expected number of particles detected on the Medipix3. Summed counts means that all hits over the measurement are summed together, this is not counting clusters. Similar count rate linearity measurements were not performed at other proton energies or with carbon ions due to limited beam time. First order corrections to this naive expectation would include a simple geometric correction and a measurement dedicated to measuring the average cluster size.

In frames with a very low number of hits, a basic visual inspection of multiple single frames shows that the average cluster size for 62.4 MeV protons is in the 3---5 pixel range. Due to the clinical fluence, almost all of the clusters are overlapping in all measurements and so clusters cannot be reliably counted.

A treatment planning system is designed to deliver dose in the 3D distribution as programmed. The way it accomplishes this is very machine dependent. This has progressed from basic methods such as rotating a radioactive sample around a patient to state-of-the-art automated systems integrated with control systems. One such modern implementation of a TPS in a particle therapy context is RayStation\textregistered, which offers a solution for PBS (pencil-beam scan) as used at MedAustron. The output of this TPS is a raster scanned pencil beam whose profile is approximately a 2D normal distribution.

\begin{figure}[tb]
    \centering
    \includegraphics[width=0.7\textwidth,keepaspectratio]{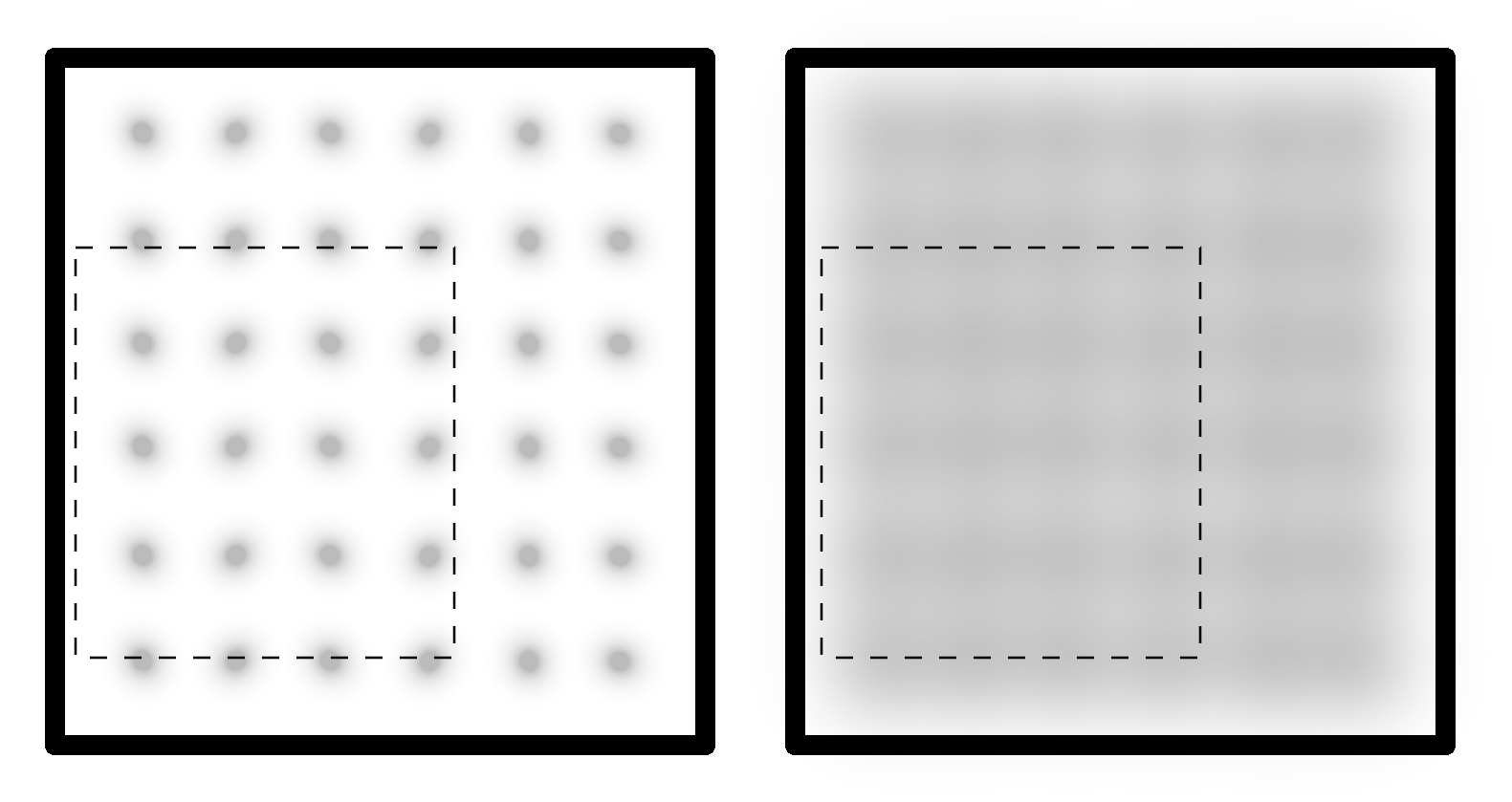}
    \caption{ An example dose distribution for a TPS (treatment planning system) using a PBS (pencil-beam scan) with a given spot FWHM (full width half maximum). A (left): large inter-spot spacing and B (right): small inter-spot spacing and many spots. The outer solid black square represents the bounding area for irradiation, the inner dashed black squares represent the detector and the grey colour shows the relative dose in arbitrary units. }
    \label{fig:20210419-MedAustron-TPS-PBS}
\end{figure}

Suppose one would like to uniformly irradiate a given area with a raster scanned pencil beam and have negligible dose outside of the designated irradiation area, it is clear that one would need to modulate the intensity of the beam over time. If we assume that the intensity of the beam can be instantaneously ramped up to the maximum and down to zero, one would expect to see a cumulative dose distribution as seen in figure \ref{fig:20210419-MedAustron-TPS-PBS}A. As one increases the number of spots and reduces the inter-spot spacing, this grid of 2D normal distributions would tend towards the distribution in figure \ref{fig:20210419-MedAustron-TPS-PBS}B which shows a decreased dose at the edges of the designated irradiation area. In this case, the irradiation zone was specified to be larger than the detector in order to approximate a uniform irradiation. This was possible because this is a non-clinical measurement which did not require minimising dose outside of the detector. A visualisation of this is shown in figure \ref{fig:20210419-MedAustron-TPS-PBS} as the dashed inner black squares representing the detectors are positioned arbitrarily within the designated irradiation area. In typical clinical settings, the designated irradiation area is determined by the treatment plan.  

Secondly, systematic errors are introduced because each proton triggers more than a single pixel as a result of the current detector front-end configuration and sensor thickness. It is known that the detector will count too many protons from previous work \cite{Yap:2019zrh} with 62.4 MeV protons, this was also observed at very low count rates during initial testing at MedAustron. A first order correction for this effect could be given by calculating mean cluster size resulting from 62.4 MeV protons in 500 $\upmu$m silicon of either simulated or measured data. The measurements would require extremely low flux in order to have average occupancy of a frame low enough to observe individual clusters. This measurement is not trivial because producing this low flux (approximately 1 kHz) of 62.4 MeV protons is completely out of the design parameters of medical accelerators. Methods used to reduce the beam intensity broaden and shift the peak energy down, therefore the energy is no longer known.

There is also an uncertainty in the summed counts from the activation of the silicon sensor, the Medipix3 chips and other surrounding material, including the readout system, aluminium cooling block and foam topped table. It was not possible to distinguish between the sources of the activation using the Medipix3 itself; alpha, beta and gamma decays were observed based on the shape of tracks. The components in the direct beam (silicon sensor, Medipix3 chips and the aluminium cooling block) are expected to be the most activated components by far. The magnitude of activation varies over time due to a combination of random radioactive decay rates, exponentially decreasing activity and the particle beam would cause increasing activation. This is expected to be proportional to dose in the sensor, chip and surrounding materials since the half-life time was in the order of days. Due to the highly mixed radiation field from various decay chains, an estimation of the dose is not made. The detector is activated to an average of 378 counts per second over the whole detector measured over 1000 seconds between two 800 MeV proton measurements half way through the day.

These effects introduce systematic uncertainties, as supported by the quality of the linear fit ($\chi_\nu^{2}$ = 4.6) for the dataset ignoring the degrader measurements. This assumes an uncertainty in the expected number of particles of 3\% due to shot-to-shot variation in the number of particles coming from the beam nozzle.

\begin{figure}[tb]
    \centering
    \includegraphics[width=0.8\textwidth,keepaspectratio]{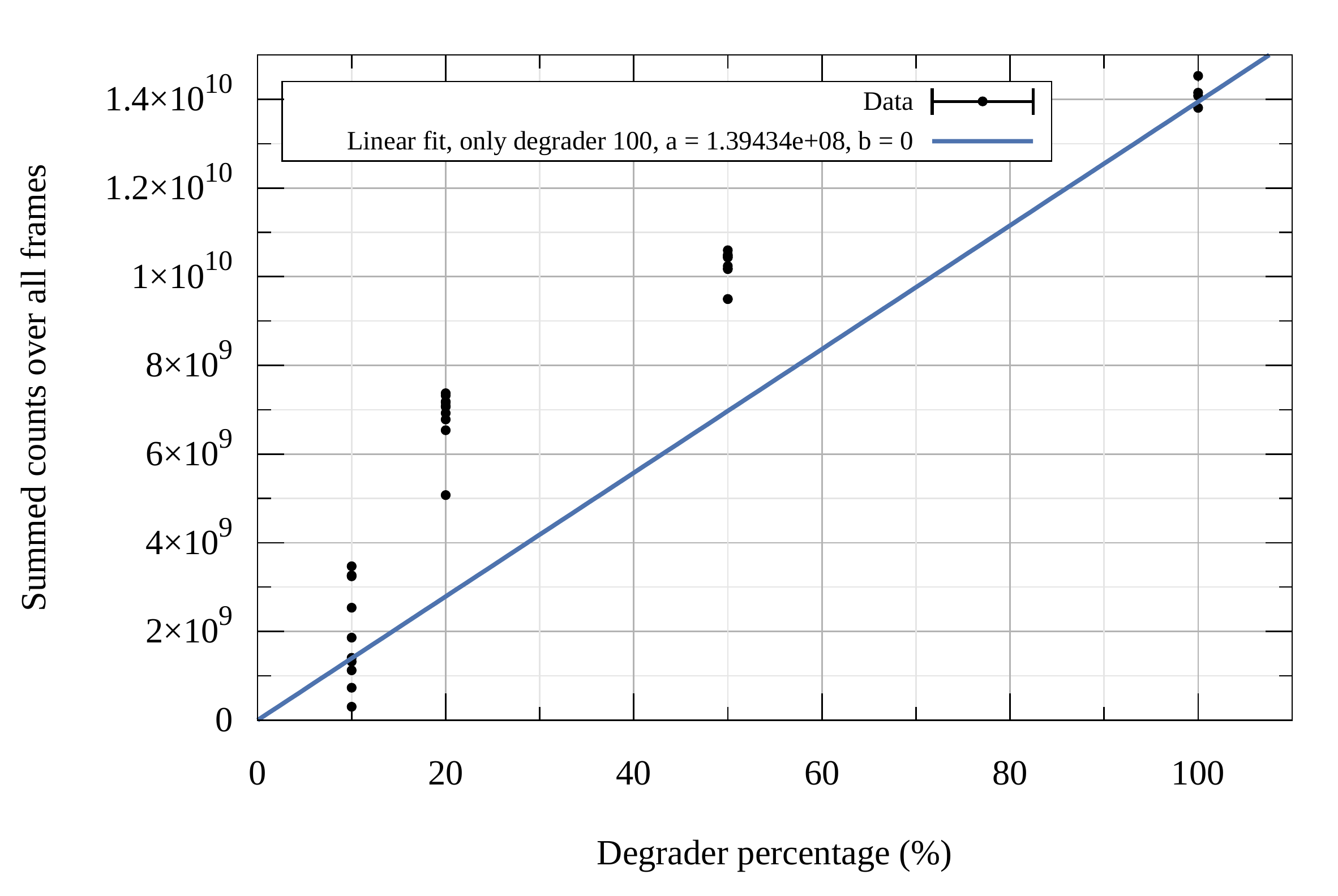}
    \caption{ The integrated counts over all frames against the degrader percentages 10, 20, 50 and 100 \% using 800 MeV protons. The linear fit only uses degrader 100 measurements and fixes the $y$-intercept to 0. No error bars are plotted because the degrader percentage is a discrete quantity and the summed counts over all frames uses the Poisson error, the square root of the counts as the error which results in errors in the order of 0.001 \% which is not visible on this scale. There are 33 data points in this plot. }
    \label{fig:p800-count-rate-linearity-all-data}
\end{figure}  

\begin{table}[htbp]
\centering
\caption{ The mean integrated counts over all frames with percentage uncertainties in the shot-to-shot intensity variation using 800 MeV protons. }
\smallskip
\begin{tabular}{|c|c|c|}
\hline
Degrader (\%) & All measurements \\
\hline
10  & $1.93\times10^{9}\,^{+80\%}_{-84\%}$ \\
20  & $6.85\times10^{9}\,^{+8\%}_{-26\%}$  \\
50  & $1.03\times10^{10}\,^{+3\%}_{-8\%}$  \\
100 & $1.41\times10^{10}\,^{+3\%}_{-2\%}$  \\ \hline                                 
\end{tabular}

\label{tab:p800-shot-to-shot-variation}
\end{table}

Figure \ref{fig:p800-count-rate-linearity-all-data} shows the count rate linearity against the degrader nominal percentage with 800 MeV protons, with and without measurements with degrader 10. Table \ref{tab:p800-shot-to-shot-variation} summarises the data from figure  \ref{fig:p800-count-rate-linearity-all-data}, showing the shot-to-shot intensity variation using 800 MeV protons with all 4 degrader plates (10, 20, 50, 100 \%). Given more measurement time, we would investigate more clinically relevant proton beam energies and also carbon ion beams. 800 MeV protons were prioritised as this was a new modality for MedAustron which was undergoing testing at the time. It is expected that this would also be linear within the uncertainties for lower energy proton beams and for carbon ion beams also.

The percentage uncertainties in superscript and subscript in Table \ref{tab:p800-shot-to-shot-variation} show the expected 2 to 3 \% uncertainty in beam intensity as measured by other devices at MedAustron during commissioning when no degraders were used. Degrader 100 \% means that no degrader was used. For the other degraders, larger uncertainties were measured. Degrader 50 consistently reduces the particle count to $\sim$75 \%. Similarly, degrader 20 reduces the particle count to $\sim$50 \%. Measurements with degrader 10 show greater relative variation, anywhere from 2---25 \% of the expected particle count is detected.
This data indicates that the various methods used to obtain lower beam currents are typically producing significantly more protons than expected based on the degrader percentage alone. It is possible that the degrader 100 measurements are suffering from saturation effects or other such losses, however this is unlikely given consistently the higher than expected counts for the degrader 10, 20 \& 50 measurements (the first three from the left) in figure \ref{fig:p62-count-rate-linearity}. A mismatch between the targeted and achieved particle count has been documented by L. Adler (Tables 6.9 \& 6.10), although these measurements may not be comparable since they are measured in different locations along the beam-line.
The large systematic uncertainties could be addressed with repeat measurements by measuring certain accelerator parameters. The total number of extracted particles can be calculated non-invasively via the differential of the main ring current transformer plus there is an active measurement of the DDS (dose delivery system) giving the exact number of particles deployed. This is calibrated in the medical energy ranges for protons: 62-252 MeV and for carbon ions: 120-400 MeV/amu. This was not measured for this work and retrieving such information from the log files is no longer possible.

If unaccounted for, this mismatch would adversely affect patient treatment because when degraders are used, one would actually be delivering more dose than intended. However, the accuracy of the degrader ratings is actually not important due to the Dose Delivery System (DDS) which is an essential component of every medical particle accelerator. The DDS is designed to measure the number of particles delivered and move to the next spot as soon as the number of particles delivered on that spot matches the requested number. Therefore, the degraders should not affect the dose delivery in actual patient treatment.

\subsection{Temporal beam intensity variation with frequency decomposition}

\begin{figure}[tb]
    \centering
    \includegraphics[width=0.7\textwidth, keepaspectratio]{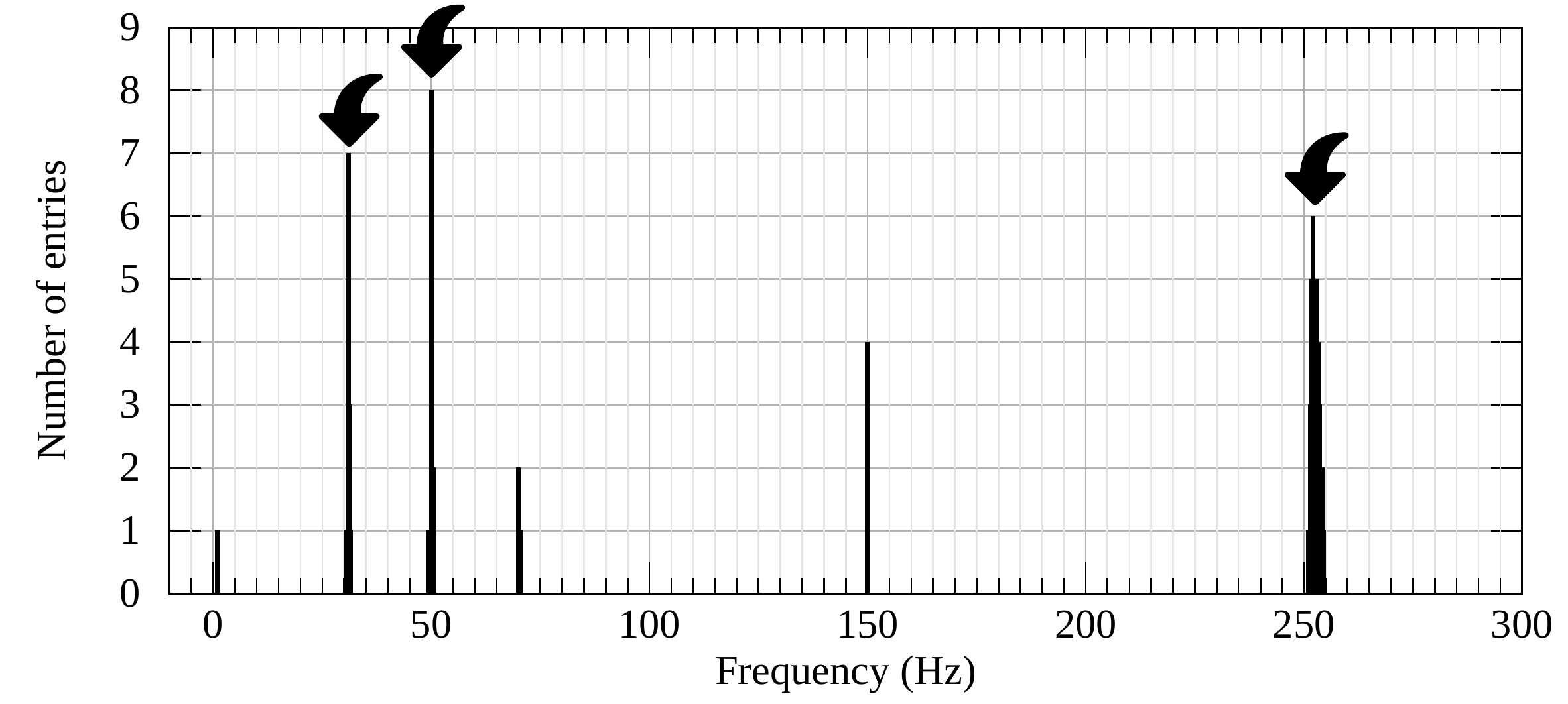}
    \caption{ A histogram of the frequency components of the intensity variation of the beams using the 1000 FPS data only. The top three in amplitude (as marked with the arrows) are 49.98 Hz with standard deviation, $\sigma=0.29$, secondly 30.55 Hz $\sigma=0.55$ and thirdly 252.51 Hz $\sigma=0.83$. The DC component (first bin of the FFT data), has been removed for visualisation purposes. }
    \label{fig:frequency-components-1000FPS-data-only}
\end{figure} 

The mean count of every raw frame over time is calculated after which a Fourier transform is applied, resulting in frequency components as shown in Figure \ref{fig:frequency-components-1000FPS-data-only}. The 49.98 Hz component is exactly the same frequency as Austrian AC mains electricity, this appears to be the most likely explanation for this frequency. The 150 Hz component is likely to be the 3rd harmonic of the mains AC frequency. The 30.55 and 252.51 Hz components are related to the spill ripples which are caused by power converter ripples and not the DDS (dose delivery system). The DDS is used for these measurements to scan over the whole detector surface.

%

\subsection{Dead/unresponsive and noisy pixels over time} \label{dead-pixels-over-time}

\begin{figure}[tb]
    \centering
    \includegraphics[width=0.7\textwidth,keepaspectratio]{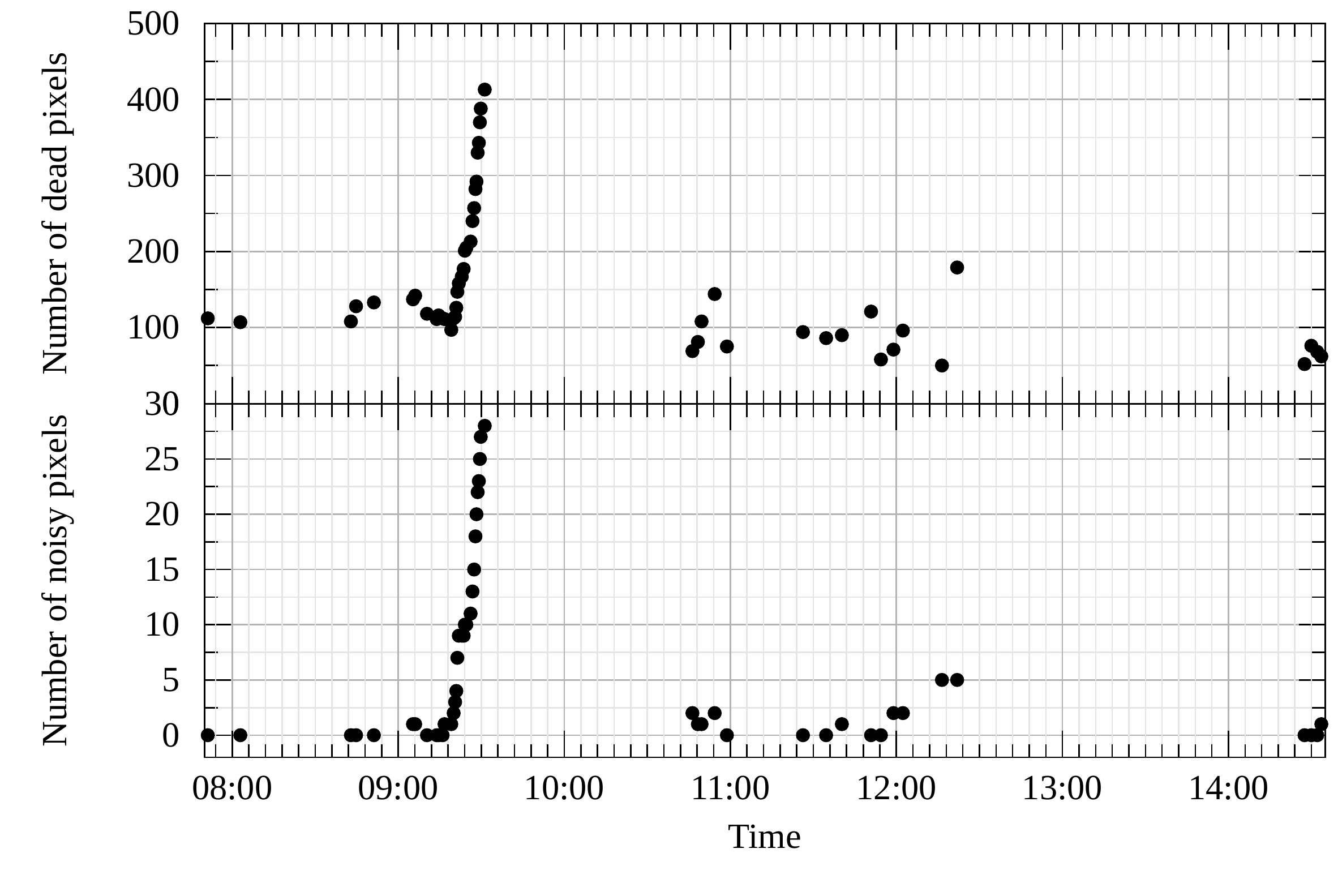}
    \caption{ The absolute number of zero counting (dead or temporarily unresponsive) pixels and noisy pixels over time (measurement start times). The percentage of dead/unresponsive pixels varies between 0.019 and 0.158 \% of the total number of pixels ($512 \times 512 = 262144$). The percentage of noisy pixels varies between 0 and 0.011 \% of the total number of pixels. }
    \label{fig:dead-and-noisy-pixels-over-time}
\end{figure}  

The total number of pixels is 262144 ($512 \times 512$ pixels). Therefore, the percentage of dead pixels over time varies between only 0.019---0.158 \% and the percentage of noisy pixels varies between 0.000---0.011 \% in figure \ref{fig:dead-and-noisy-pixels-over-time}. A slow and precise equalisation (a procedure to flatten pixel noise baseline levels over the chip \cite{Rinkel2015}) was started at 09:30 and finished an hour after. This procedure was run to correct for perceived damage as well as an increased number of bad pixels, i.e. noisy, dead, unresponsive pixels. There was a significant reduction in bad pixels compared to pre-irradiation levels, and these numbers were relatively stable for subsequent runs. There is a strong correlation between bad pixels for the high flux region between 09:15 and 09:30 whereas the other points show low correlation.

It is possible that during this high flux region in time, the Si-SiO$_2$ interface between the pixel implants and the sensor bulk accumulated charge to the point where the affected pixel's noise baseline was shifted out of range. If a pixel's noise baseline is shifted out of range, it will either never respond or would be noisy at that threshold. This effect is observed by measuring increased leakage current during relatively high intensity x-ray irradiation resulting in higher noise baselines over time. The leakage current decreases back to the pre-irradiation levels consistently.

Given that an equalisation and some time with no radiation fixed this issue and similar behaviour is observed at a much slower rate with x-rays, this hypothesis appears to be consistent. More studies would be relevant to probe the exact underlying mechanism.

\subsection{Carbon ion energy dependency on total counts} \label{carbon-ion-energy-dependence-on-total-counts}

Carbon ion spills lasting 22 to 25 seconds at 120, 260 and 400 MeV/amu were measured in order to verify the expectation that carbon ion energy should be inversely proportional to the total counts recorded. As the carbon ion energy increases, the probability of interaction per unit length in the silicon decreases. As table \ref{tab:carbon-energy-counts} shows, the carbon ion energy is inversely proportional to the total number of counts. This dependency was measured to be linear within this region and the number of carbon ions requested was kept constant.

\begin{table}[tb]
\centering
\caption{ The carbon ion energy dependency on the total number of counts per spill. A linear fit on this data results in a slope of $-1.26 \times 10^{7}$, an intercept of $1.27 \times 10^{10}$, with $R^{2}=0.978$. }
\smallskip
\begin{tabular}{|c|c|}
\hline
Energy (MeV/amu) & Total number of counts \\
\hline
120            & $1.10 \times 10^{10}$  \\
260            & $9.74 \times 10^{9}$   \\
400            & $7.53 \times 10^{9}$   \\  \hline           
\end{tabular}
\label{tab:carbon-energy-counts}
\end{table}  

\subsection{Radiation damage}

\subsubsection{During proton and carbon ion measurements}

\begin{figure}[tb]
    \centering
    \includegraphics[width=0.7\textwidth,keepaspectratio]{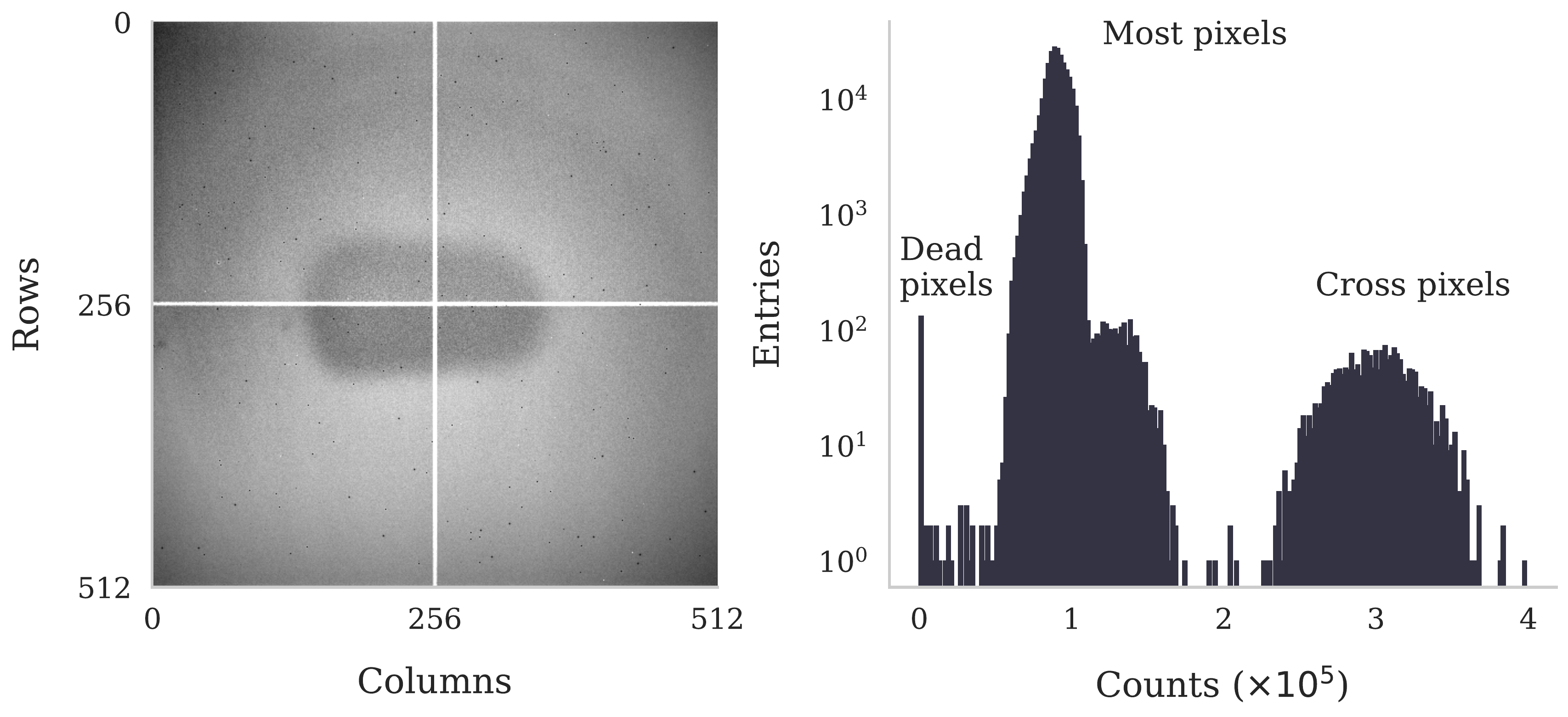}
    \caption{Left: An integrated image of all 4 chips over all frames with 62.4 MeV protons scanned over the surface using the Treatment Planning System, a uniform exposure was intended. The darker oval shaped region in the centre shows the area suspected of radiation damage. The view is optimised for non-cross pixels which is why the cross pixels are all white. The x-axis shows detector columns and the y-axis shows detector rows. Right: A histogram of the integrated image, showing the regions of interests: dead pixels, most pixels and the cross pixels.}
    \label{fig:radiation-damage-during-measurements}
\end{figure}  

Figure \ref{fig:radiation-damage-during-measurements} shows an approximate 10\% response decrease in the counts over the selected region of interest. Ideally, the detector would have been uniformly irradiated via requests to the DDS, which scans over the surface using a number of spots at a particular target spot weight. Variations in the extracted number of particles per spot and between spots introduce uncertainty to the homogeneity of the delivered particle distribution, these effects were not calculated. 

\subsubsection{X-ray imaging 37 days after the proton and carbon ion measurements}

The aim of this measurement was to find evidence of radiation damage and if it was consistent with supposedly uniform proton irradiation. This is achieved by using a cone beam x-ray tube to produce a relatively homogeneous radiation field. X-rays are the lowest energy, individually detectable particles with the Medipix3, can be produced at high rates ($>10^{11}$ / s) with common x-ray tubes and do not damage the detector at this flux. X-rays are therefore appropriate for investigating the homogeneity of the detector response over the surface after irradiation with particles causing nuclear interactions in the silicon sensor such as protons and carbon ions in the MeV range and above.

Relevant parameters: x-ray tube peak voltage 50 kVp, tube current 0.92 mA, 5 minute exposure and the detector was 15 cm from the tube exit window. These parameters were chosen in order to produce a homogeneous field with a very high number of x-rays resulting in a very low statistical uncertainty of $<$ 0.001\%. The tube peak voltage and the currents are the maximum possible for this Jupiter 5000 Series x-ray tube \cite{oxfordInstruments}.

\begin{figure}[tb]
    \centering
    \includegraphics[width=0.7\textwidth,keepaspectratio]{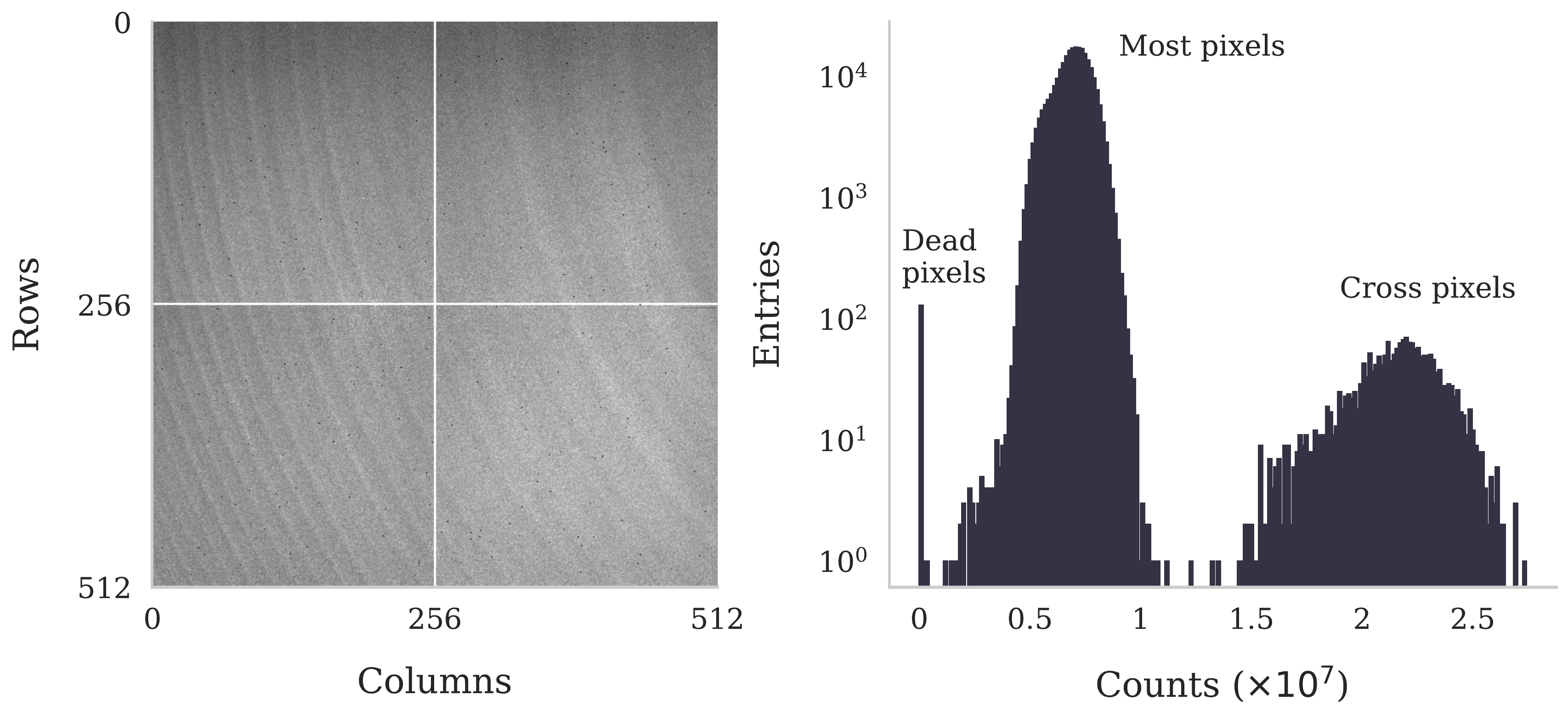}
    \caption{Left: A raw x-ray image of all 4 chips over 5 minutes where the x-axis shows detector columns and the y-axis shows detector rows. The view is optimised for the majority of the pixels, not the cross pixels. Right: A histogram of the integrated image, showing the regions of interests: dead pixels, most pixels and the cross pixels.}
    \label{fig:after-radiation-damage}
\end{figure}  

One can observe from figure \ref{fig:after-radiation-damage} that there is not a decrease in response in the centre of the image consistent with the uniform proton irradiation image \ref{fig:radiation-damage-during-measurements}. This implies that either the proton response is different from the x-ray response, the supposed radiation damage observed during irradiation, as seen in \ref{fig:radiation-damage-during-measurements}, has annealed or a combination of both. Damage is expected there due to the proton beam being fixed on the central area in the first part of the day.

As for detector uniformity, 9.9 \% variation in counts across the detector surface is typical in this configuration and does not indicate radiation damage. 95 \% of counts are within 19 \% of the mean, averaged over the 4 chips, excluding the cross. No response variation is observed in the centre.

Regarding the two patterns visible in figure \ref{fig:after-radiation-damage}, the wave-like pattern across the 4 chips is known to be due to doping concentration variation in the p-on-n silicon sensor during the crystal growth \cite{lim_forster_zhang_holtkamp_schubert_cuevas_macdonald_2013}. P type doping with boron is used for the implant and phosphorus is used for N type doping for the n bulk. A single silicon sensor cut from a single wafer is bump bonded to 4 Medipix3 chips in a 2x2 grid in this configuration. This explains why the waves are continuous across all the chips. The vertical gradient is due to the detector being close enough to the x-ray tube that the cone beam has significant intensity variation.

The number of dead/unresponsive pixels was 112 (0.043 \%) at the start of the proton and carbon ion measurements and is 123 (0.047 \%) in this x-ray test. There are only 0.0004 \% more dead/unresponsive pixels than before any proton and carbon ion irradiation. There is variation in this number as shown in section \ref{dead-pixels-over-time}. Given that the variation in that number is much larger than the difference here, it is not expected that this difference is significant.

An average increase of 2.5 DAC (digital-to-analogue) units is observed; the mean of the noise of the chips is 1 \% more than before the measurement. Given that the temperature was not monitored, this is within the expected variation and is therefore not a conclusive indication of increased chip noise. Simulations of the Medipix3 chip response to temperature were done during the design process \cite[5.4.5 \& Appendix III]{Ballabriga2009} where Ballabriga simulates that the shaper output signal should vary by 0.16 \%/$^{\circ}$C with nominal settings. The shaper output signal magnitude is directly proportional to the aforementioned DAC values. This would imply a temperature difference of 6.25 $^{\circ}$C which is consistent with expectations.

In conclusion, no significant increase in dead/unresponsive pixels is observed. The cause of the wavy pattern is well known. The total variation in response over the detector is in the normal range. No significant increase in chip noise is observed. No reasonable estimation of detector lifetime can be calculated from this. Studies have investigated the radiation hardness of the Medipix3 chip with x-rays and neutrons, for x-rays the chip was still operational after 460 MRad at a high dose rate of 3.5 kGy/s \cite{Plackett2009} and for neutrons, the chips tolerated 1 MeV neutron equivalent dose of $5 \times 10^{14}$cm$^{-2}$ \cite{Akiba2016}. This is far in excess of typical requirements for space grade radiation hardness of ~300 kRad and more in line with LHC inner tracker requirements of ~300 MRad. Simulations are required to calculate the dose in the detector, in order to do this, one would need to know the ratios of elements in order to estimate the relation between particles of a particular energy and the absorbed dose in the chip, this was not in the scope of this analysis.

Additional tests at clinically relevant dose rates ($>$ 1 Gy/s in water) with proton and carbon ion beams are needed to determine the lifetime of the detector in this radiation environment. Due to the limited beam time and detector supply, this is not in the scope of this study.

Based on this study, it is expected that the Medipix3 chips would be used as beam profile monitors for relatively infrequent quality assurance (QA) measurements. Additionally, these chips are relatively effective at verifying the accuracy and precision of the DDS. This was the first measurement to simultaneously show the spatial and temporal distribution of delivered protons and carbon ions at MedAustron; the Medipix3 fills that niche.

Many different detector geometries could be suitable for this application depending on the exact intended use, for example, one could exclusively measure the edges of the beam so the system would not disrupt the beam and could run indefinitely. The other extreme would be to use a retractable large area detector for occasional quality assurance measurements.

Scaling of Medipix3 based detectors is possible with TSV (Through Silicon Via) technology which enables N$\times$N scaling, subject to sensor wafer size primarily \cite{tsv1, tsv2, tsv3} with a 0.8 mm non-active area, the periphery. The Medipix4 is being designed to further improve on this by eliminating the non-active area, enabling 100 \% active detection area.

\section{Conclusions}

The Medipix3 chip with a 500 $\upmu$m silicon sensor has been used for a series of measurements using high energy protons and carbon ions with a wide range of particle flux and energies. Protons with energies of 62.4, 148, 252 and 800 MeV were used at flux rates between $10^{4}$ and $10^{8}$.
120, 260, 400 MeV/amu carbon ions were used at flux rates varying between $10^{7}$ and $10^{8}$ carbon ions per second impinging on the detector surface.

The temporal beam intensity variations were decomposed into frequency components showing several peaks including Austrian mains frequency and two others which are related to the spill ripples in the synchrotron. None of these degrade the patient treatment due to the design of the Dose Delivery System (DDS).

During the period of highest flux, the number of zero counting and noisy pixels increased rapidly and were correlated. After running a software procedure to equalise the pixel response over the matrix, the number of zero counting and noisy pixels returned to approximately pre-irradiation levels. Further studies would be relevant to investigate this effect.

There is evidence that the Medipix3 can be used as a beam instrumentation device. It shows good count rate linearity with 62.4 MeV protons over the full flux range available, reliable performance at 1000 FPS and is sensitive to single particles. Proton and carbon ion beams have been measured at the full energy range, respectively 62.4 to 800 MeV and 120 to 400 MeV/amu. No conclusive evidence of radiation damage was observed, further measurements are necessary to determine detector lifetime. The Medipix3 front-end settings (DACs) could be optimised and tested from the default low energy x-ray ($<$ 30 keV) configuration with more beam-time.

\acknowledgments
This work is part of the research programme of the Foundation for Fundamental Research on Matter (FOM), which is part of the Netherlands Organisation for Scientific Research (NWO). It was carried out at the National Institute for Subatomic Physics (Nikhef) in Amsterdam, the Netherlands.

This project has received funding from the European Union's Horizon 2020 research and innovation programme under the Marie Skłodowska-Curie grant agreement No. 675265 - OMA (Optimization of Medical Accelerators).

\appendix

\section{Measurement overview}
An overview of all measurements is displayed in tables \ref{tab:overview} and \ref{tab:overview2}.

\begin{table}[htbp]
\centering
\caption{ Measurements overview, 1 of 2. Spot weight has the units of numbers of particles specified. }
\smallskip
\begin{tabular}{|c|c|c|c|c|c|}
\hline
Run & Sub-run & Particle & Energy (MeV) & Degrader (\%) & Spot weight      \\
\hline
1   &         & Proton        & 800          & 20            & N/A         \\
2   &         & Proton        & 800          & 20            & N/A         \\
3   &         & Proton        & 800          & 20            & N/A         \\
4   & Test    & Proton        & 800          & 10            & N/A         \\
    & 1       & Proton        & 800          & 10            & N/A         \\
    & 2       & Proton        & 800          & 10            & N/A         \\
5   & 1       & Proton        & 800          & 10            & N/A         \\
    & 2       & Proton        & 800          & 10            & N/A         \\
    & 5       & Proton        & 800          & 10            & N/A         \\
    & 7       & Proton        & 800          & 10            & N/A         \\
    & 8       & Proton        & 800          & 10            & N/A         \\
    & 9       & Proton        & 800          & 10            & N/A         \\
    & 10      & Proton        & 800          & 10            & N/A         \\
6   & 1       & Proton        & 800          & 20            & N/A         \\
    & 2       & Proton        & 800          & 20            & N/A         \\
    & 3       & Proton        & 800          & 20            & N/A         \\
    & 4       & Proton        & 800          & 20            & N/A         \\
    & 5       & Proton        & 800          & 20            & N/A         \\
    & 6       & Proton        & 800          & 20            & N/A         \\
    & 7       & Proton        & 800          & 20            & N/A         \\
    & 8       & Proton        & 800          & 20            & N/A         \\
    & 9       & Proton        & 800          & 20            & N/A         \\
    & 10      & Proton        & 800          & 20            & N/A         \\ \hline
\end{tabular}
\label{tab:overview}
\end{table}

\begin{table}[htbp]
\centering
\caption{ Measurements overview, 2 of 2. Spot weight has the units of numbers of particles specified. }
\smallskip
\begin{tabular}{|c|c|c|c|c|c|}
\hline
Run & Sub-run & Particle & Energy (MeV) & Degrader (\%) & Spot weight      \\
\hline
7   & 1       & Proton        & 800          & 50            & N/A         \\
    & 2       & Proton        & 800          & 50            & N/A         \\
    & 3       & Proton        & 800          & 50            & N/A         \\
    & 4       & Proton        & 800          & 50            & N/A         \\
    & 5       & Proton        & 800          & 50            & N/A         \\
    & 6       & Proton        & 800          & 50            & N/A         \\
    & 7       & Proton        & 800          & 50            & N/A         \\
    & 8       & Proton        & 800          & 50            & N/A         \\
    & 9       & Proton        & 800          & 50            & N/A         \\
    & 10      & Proton        & 800          & 50            & N/A         \\
    & 11      & Background    & 800          & N/A           & N/A         \\
8   & 0       & Proton        & 800          & 100           & N/A         \\
    & 2       & Proton        & 800          & 100           & N/A         \\
    & 3       & Proton        & 800          & 100           & N/A         \\
    & 4       & Proton        & 800          & 100           & N/A         \\
9   &         & Proton        & 62           & 20            & $5 \times 10^6$    \\
10  &         & Proton        & 148          & 20            & $1 \times 10^7$    \\
11  &         & Proton        & 252          & 20            & $1 \times 10^7$    \\
12  &         & Proton        & 62           & 10            & $1 \times 10^6$    \\
13  &         & Proton        & 62           & 100           & $1 \times 10^8$    \\
14  &         & Proton        & 62           & 100           & $5 \times 10^7$    \\
15  &         & Proton        & 62           & 50            & $1 \times 10^7$    \\
16  &         & Proton        & 62           & 100           & $1 \times 10^9$    \\
17  &         & Proton        & 62           & 100           & $5 \times 10^8$    \\
18  &         & Carbon        & 120          & 20            & N/A         \\
19  &         & Carbon        & 400          & 20            & N/A         \\
20  &         & Carbon        & 260          & 20            & N/A         \\
21  &         & Carbon        & 120          & 100           & N/A         \\ \hline
\end{tabular}
\label{tab:overview2}
\end{table}

\end{document}